\documentclass[fleqn,usenatbib]{mnras}
\usepackage[T1]{fontenc}
\usepackage{ae,aecompl}
\usepackage{graphicx}% Including figure files                                                       
\usepackage{amsmath}% Advanced maths commands                                                       
\usepackage{amssymb}% Extra maths symbols                                                           
\usepackage{txfonts}

\title[A galaxy morphology classification grid]
{A galaxy classification grid that better recognises early-type galaxy morphology}

\author[A.\ W.\ Graham]{
Alister W.\ Graham\thanks{E-mail: AGraham@swin.edu.au}\\
Centre for Astrophysics and Supercomputing, Swinburne 
University of Technology, Hawthorn, Victoria 3122, Australia.
}

\date{Accepted 2019 June 7. Received 2019 June 7; in original form 2019 April 26}

\pubyear{2019}

\begin{document}
\label{firstpage}
\pagerange{\pageref{firstpage}--\pageref{lastpage}}
\maketitle

\begin{abstract}

A modified galaxy classification scheme for local galaxies is presented.  It
builds upon the Aitken-Jeans nebula sequence, by expanding the Jeans-Hubble
tuning fork diagram, which itself contained key ingredients from Curtis and
Reynolds.  The two-dimensional grid of galaxy morphological types presented
here, with elements from de~Vaucouleurs' three-dimensional classification
volume, has an increased emphasis on the often overlooked bars and continua of
disc sizes in early-type galaxies --- features not fully captured by past
tuning forks, tridents, or combs. The grid encompasses nuclear discs in
elliptical (E) galaxies, intermediate-scale discs in ellicular (ES) galaxies,
and large-scale discs in lenticular (S0) galaxies, while the E4-E7 class is
made redundant given that these galaxies are lenticular galaxies.
Today, these structures continue to be neglected, or surprise researchers,
perhaps partly due to our indoctrination to galaxy morphology through the
tuning fork diagram.  To better understand the current and proposed
classification schemes --- whose origins reside in solar/planetary formation
models --- a holistic overview is given.  This provides due credit to some of
the lesser known pioneers, presents some rationale for the grid, and reveals
the incremental nature of, and some of the lesser known connections in, the
field of galaxy morphology.

\end{abstract}


\maketitle
 
\section{Introduction}  
\label{Sec:Intro}  

Clues to the past evolution of galaxies can be found encoded within their
morphologies.  The principal structures are the spheroid and disc, and the bar
and spiral arms within the disc.  A fundamental galaxy classification scheme
should, therefore, capture these elements at some level, and may help to
reveal links between observations and physical processes.  For example, in the
mid-1900s, an often heralded justification / success of the
Jeans-Lundmark-Hubble galaxy sequence (Jeans 1919a; Lundmark 1925, 1927;
Hubble 1926) was how it tracked the mean age of the host galaxies' stellar
population.

Although the Jeans-Hubble tuning fork diagram (Jeans 1928, Hubble 1936)
provided minimal information about the early-type galaxies (ETGs), the greater
detail regarding the late-type galaxies (LTGs) was an invitation that
theorists and modellers responded to (e.g.\ Roberts et al.\ 1975; Feitzinger
\& Schmidt-Kaler 1980; Toomre 1981, and references therein; Haass et
al.\ 1982).  In contrast, the long thin handle of the ``tuning fork'' likely
hindered an awareness as to the prevalence of substructures, such as discs and
bars, within the ETG population (E, ES\footnote{Bridging the elliptical
  (largely discless, except for an optional nuclear disc) and lenticular galaxies (with large-scale discs
  dominating the light at large radii), Liller (1966) identified the ES galaxy type with
  intermediate-scale discs (e.g.\ Capaccioli \& Vietri 1988; Michard \& Simien
  1988; Nieto et al.\ 1988, 1991; Simien \& Michard 1990; Michard \& Marchal
  1993).  Following Liller, who concatenated the letters E and S0 to give the
  ES galaxy type, Graham et al.\ (2016) concatenated the words elliptical and
  lenticular, to give the ``ellicular'' galaxy name.}, and S0).  This
situation was partly remedied by the Hubble-influenced galaxy classification
schemes of de Vaucouleurs (1959a,b) and Sandage (1961), but the level of
complexity in their schemes may have inhibited a more wide-spread uptake.

The new galaxy morphology classification grid shown herein, stemmed from the
recent analyses of several ETGs with intermediate-scale discs. This started
with an investigation into the remarkably boxy, dwarf ETG known as LEDA~074886 (Graham
et al.\ 2012). This galaxy harbours an edge-on, intermediate-scale disc --- whose rotation
was revealed using the Keck Telescope --- and it was unclear where such a 
galaxy resides in the galaxy classification schemes.  It was similarly 
problematic trying to place the dwarf ES galaxy CG~611 (Graham et al.\ 2017,
see their section~3.3.2) into the spin-ellipticity diagram of slow and fast
rotators (Emsellem et al.\ 2007, 2011).  In this galaxy, the
intermediate-scale disc appears face-on, revealing the presence of a bar.
Such fully-embedded discs are not confined to dwarf ETGs. For example, Section 3.3 of Graham
et al.\ (2016) reports on the massive ES galaxy NGC~1271 in the Perseus cluster. Additional recent
examples of ordinary ETGs with intermediate-scale discs can be seen in 
Savorgnan \& Graham (2016b) and Sahu et al.\ (2019).

The ETGs have long been a misclassified, and consequently
misunderstood, population.  Indeed, the discs of lenticular galaxies were not
originally recognised as such, but rather they were regarded as demarking the
``fundamental plane'' of lentil-shaped galaxies.  Furthermore, unless
these discs were orientated rather edge-on to our line-of-sight, or contained
circular dust lanes/rings, then they usually went undetected.  As noted by
Capaccioli (1990), most ETGs contain a large-scale disc, and many of these
discs contain a bar, while some also contain ansae (Martinez-Valpuesta et
al.\ 2007) at the ends of the bar, and rings.  Studying galaxy
images, Rix \& White (1990) reported that almost all non-boxy\footnote{The term
  ``boxy'' refers to galaxies whose isophotal shapes are slightly more
  rectangular than elliptical.}  ``elliptical'' galaxies could have discs, and
studying the kinematics of bright elliptical galaxies, Graham et al.\ (1998)
highlighted that the actual number of dynamically hot\footnote{``Dynamically
  hot'' refers to a system where random (stellar) motion dominates over
  ordered (stellar) motion, i.e.\ where a galaxy's velocity dispersion
  dominates over the rotational support.}  stellar systems is much lower than
previously thought.  Weak bars have also tended to be overlooked, as noted by
Guti\'errez et al.\ (2011) and found by Sahu et al.\ (2019).  Some of the
general ongoing confusion surrounding the ETG population, arising from their
treatment as discless, one-component systems, may, in part, originate from
their inadequate representation in the tuning fork diagram that we were all
undoubtedly introduced to through countless good textbooks and online
resources.

In order to develop or modify a galaxy classification scheme that more fully
captures the morphology of ETGs, the origins of, and many subsequent
variations to, the tuning fork diagram are reviewed in Sections~\ref{Sec_Hist}
and \ref{Sec_Mods}, respectively.  What becomes apparent is the incremental
nature of progress, and the continual re-working of particular themes.  In
that regard, this paper is no different.  Section~\ref{Sec_Grid} presents a
diagram, with a grid structure that better represents the ETGs.  The new
feature of this grid is the inclusion of Martha Liller's ES galaxies,
capturing the range of disc sizes in ETGs.  Although known for half a century
(Liller 1966; di Tullio 1978), the intermediate-scale discs of ES galaxies
have repeatedly been overlooked.

It is hoped that the grid will help facilitate a greater awareness of the
primary structures in ETGs, which are still not immediately recognised in some
galaxy research today.  Section~\ref{Sec_Disc} includes a discussion of
possible evolutionary paths for ETGs within the grid, points out connections
with kinematic classifications, and provides an outlook into large automated
surveys in which the capacity of human classifiers to assign a morphological
type has been surpassed.

\section{Historical Briefing}\label{Sec_Hist}

In an effort to distill the patterns seen in the nebula catalogs of La Caille
(1755) and Messier (1781), and of course the larger imaging campaigns of both
the Herschel family (1786, 1864) and Parsons\footnote{William Parsons, the
  3$^{\rm rd}$ Earl of Rosse, discovered the spiral nebulae (Rosse 1850a,b).}
(1878) --- which led to the New General Catalogue (NGC: Dreyer 1888) and the
supplementary Index Catalogue (IC: Dreyer 1895) --- new classification schemes
were explored by Wolf (1908), Knox Shaw (1915), Curtis (1918), and others.
Collectively, this contributed towards the Jeans (1919a,b) evolution sequence
for galaxies, which drew direct analogy from the Laplace (1796, 1799-1825)
model\footnote{Prentice (1978, 1984; see also Prentice \& Dyt 2003) developed
  the modern Laplacian theory.  Today, there is competition from the solar
  nebular disc model (Safronov 1969).} of planetary formation, in which a
cooling and thus contracting\footnote{The collapse is described by the Jeans
  (1902) instability that was advanced to explore the ``nebular hypothesis''
  of Swedenborg (1734), Kant (1755), and Laplace (1796, 1799-1825).} nebula
forms a rotationally-induced equatorial bulge\footnote{See also Gott (1975),
  Freeman (1975), and Larson (1975) for related models.}, 
 before throwing off rings\footnote{Only rings, not spirals,
  are seen in the early, nebulae catalogue drawings (e.g.\ Herschel 1833, his
  Figure~25).  As such, the Laplace (1825) model involves the generation of
  rings rather than spirals.} of matter, into the equatorial plane, out of which
planets subsequently condense.

Following ongoing works pertaining to solar system formation (e.g.\ Tisserand
1889-1896; Keeler 1900, his p.348; Chamberlin 1901; Moulton 1905; and Aitken
1906, see his p.118-119), Jeans (1919a,b) embraced the idea that as a nebula
ages, it will form a lenticular shape due to its increased rotation.  He
similarly embraced the Aitken (1906) modification to the Laplace model, in
which an external gravitational force from a passing nebula will invoke the
formation of 
symmetric spiral arms --- often modelled as ``equiangular'' spirals,
i.e.\ logarithmic spirals (von der Pahlen 1911; Groot 1925; Davis et
al.\ 2017) --- rather than rings.  It was speculated that this external
gravitational perturbation would result in a system-wide tide --- akin to the
Moon-induced tides on either side of the Earth --- sufficient for the
tidal theory of Roche (1850) to generate near- and far-side protrusions from
the nebula that eventually lead to spiral arms because of the system's
rotation (Aitken 1906; see also Alexander 1852).  These arms would grow out of
the (shrinking) nebula's material, and move outward with time, becoming more
open and eventually producing (gravitational) condensations which, in the galaxy scenario, are
not planets but star clusters.\footnote{In the early 1900s, before the
  ``island universe'' nature of the distant ``nebulae'' were known, some
  astronomers thought that the spiral nebulae, which we now know are spiral
  galaxies, were solar systems in formation (e.g.\ Sutherland 1911, but see
  Schwarzschild 1913).}  The Laplace-Aitken model for solar nebulae is the
basis of the Jeans (galaxy) ``nebula hypothesis'' for a sequence from smooth
early-type (young) elliptically-shaped nebulae progressing toward lenticular
nebulae, then early-type spirals with prominent lenticular-shaped bulges, to
late-type (old) spiral nebulae, which have small bulges, are increasingly
complex in appearance, and have open spiral arms and increasingly noticeable
condensations of stars in these arms.  This galaxy evolution sequence was
popularly discussed in the early 1920s (e.g.\ Reynolds 1921, 1925).  The
spheroid or central bulge, and the spiral arms, are primary elements of this
scheme.  Reynolds (1925, see his p.1016) also remarks that a limited number of
edge-on lenticular nebula, e.g.\ NGC~5868 and NGC~2859, were already known at
that time, bridging the elliptical and spiral nebulae.

\begin{figure*}
\begin{center}
\includegraphics[trim=5.0cm 0cm 5.2cm 0cm, width=0.25\textwidth, angle=-90]{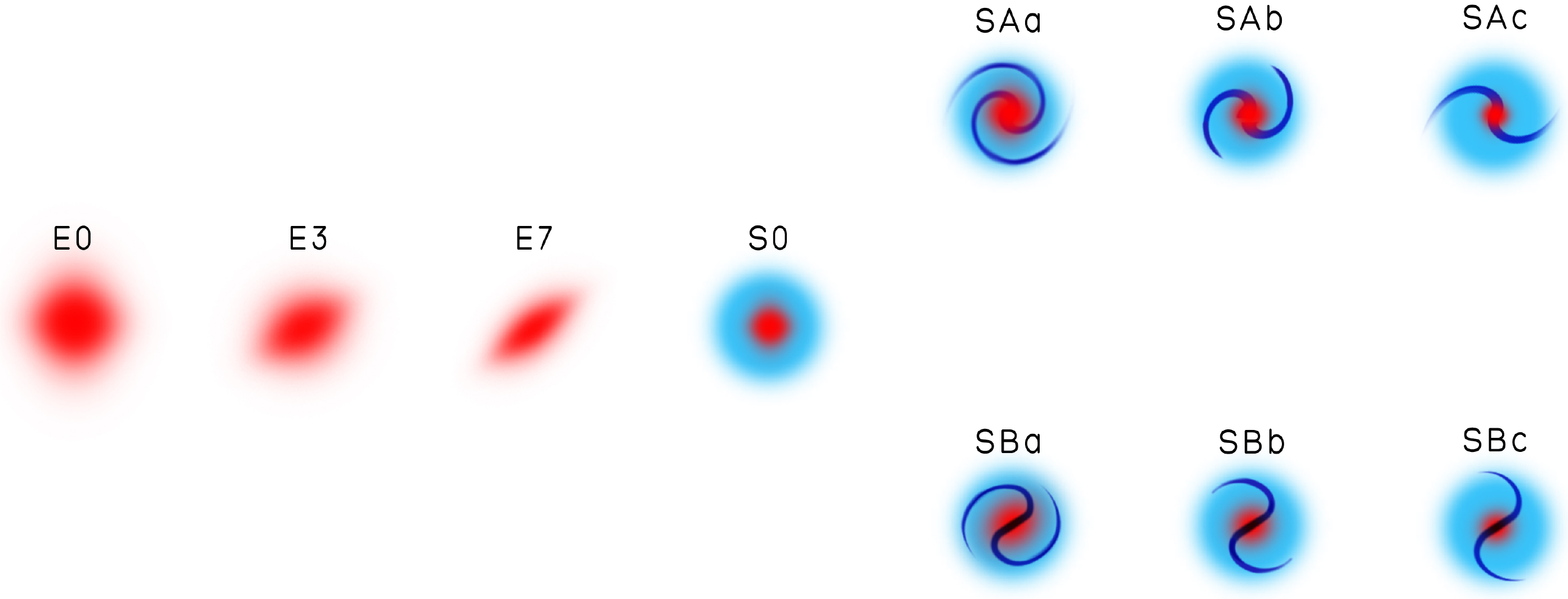}
\caption{Jeans-Hubble tuning fork (Jeans 1928; Hubble 1936), including the S0
  galaxy morphological type theorised in Jeans (1919a), identified in Reynolds
  (1925, see his p.1016), and later added by Hubble (1936).} 
\label{fig:tuning_fork}
\end{center}
\end{figure*}

Supplanting the classification proposed by Hubble (1922), 
Hubble (1926)\footnote{Hubble (1926) built on a manuscript that Edwin Powell Hubble sent to
  Vesto Melvin Slipher in 1923 for circulation to the members of Commission 28, on
  Nebulae, at the upcoming second I.A.U.\ meeting to be held in 1925.}  adopted this broad
sequence from Jeans (1919a,b), while also incorporating the spiral sequence
from Reynolds (1920).  However, by the mid-1920s most papers were dropping the idea
of a temporal sequence, and Hubble (1926, footnote 1 of p.326) also now
distanced himself from the temporal connotation which he had initially built
into his classification scheme through his use of the words ``early''
and ``late''.\footnote{In Hubble's initial manuscript, 
 which he sent to Slipher in 1923, he wrote that
  ``there is some justification in considering the elliptical nebulae as
  representing an earlier stage of evolution'', and in 1923 he also now listed them
  {\it before} the spiral nebulae (cf.\ Hubble 1922, his p.168; see also the detailed
  review by Hart \& Berendzen 1971, their p.114).} While Hubble (1926) now embraced only the
morphological sequence of Jeans and not the evolutionary
sequence\footnote{Block \& Freeman (2009) discuss how Hubble sought out
  Reynolds help, apparently prior to 1926, to develop a classification scheme
  that would be comparable to the Jeans' evolution
  sequence. Jeans (1931) subtly notes that Hubble succeeded.}, in mimicking
Jeans, Hubble (1926) had, somewhat ambiguously, written that ``the arms appear
to build up at the expense of the nuclear regions and unwind as they grow; in
the end, the arms are wide open [highly resolved] and the nuclei
inconspicuous''.

The first criteria used by Hubble (1926) was the relative size of the
unresolved nuclear region, i.e.\ the bulge or spheroid.  Then followed the
ellipticity in the case of the ETGs, and for the LTGs it was the degree of
resolution in the spiral arms, and the extent to which they are unwound.
Interestingly, Lundmark (1925, 1927) used concentration, rather than
ellipticity, to classify the then-called globular nebulae, i.e.\ the elliptical
objects, in his scheme.  As such, while Hubble (1926) focussed
on the progression of forms (ellipticity, spiral arms), Lundmark had
inadvertantly hit upon what we now know is a concentration-mass sequence among
the ETGs\footnote{Lundmark (1925) also used concentration to separate the nebulae 
  within the spiral class.} (e.g.\ Davies et al.\ 1988; Caon et al.\ 1993).
Shapley (1927) and Shapley \& Ames (1929) would subsequently use both
concentration and elongation of form in their Virgo-Coma galaxy catalog. 

Hubble (1926) additionally included a parallel sequence for the non-barred
(Normal) and barred ($\phi$-type) spiral nebulae that had been pointed
out by Knox-Shaw (1915) and Curtis (1918, his page~12).  While undoubtedly
something of a cut-and-paste job, Hubble's classification scheme based on his
summary of appropriate pieces of information --- which is often how science
progresses --- resulted in a concise and much used synopsis of extra-galactic
anatomy.  The classification summarised in Hubble (1926) included Elliptical
(E0-E7), Normal, i.e.\ non-barred, Spiral (Sa, Sb, Sc), Barred
Spiral (SBa, SBb, SBc), and Irregular (Irr) nebulae. 

Reynolds (1927) called Hubble's attention to pre-existing and rather similar
classification schemes which were not cited in Hubble (1926).  This followed
Hubble (1926, footnote~2 of p.323) which called-out Lundmark (1925, see his
p.867) for having presented a ``practically identical'' classification scheme
but having not cited Hubble.  Lundmark had, however, established his
classification scheme by 1922, before Hubble sent his initial manuscript to
Slipher in 1923 (see Teerikorpi 1989, and page~24 in Lundmark 1927).  Reynolds
was also likely motivated to act as the ethical police given that Hubble
(1923) had found that Reynolds' (1913, 1916, 1920) model for the bulges of
spiral nebulae, also described elliptical nebulae. Yet Hubble (1923) gave no
credit to Reynolds, whose model subsequently became known as the Hubble law,
or occasionally the Hubble-Reynolds law, not to be confused with the
Hubble-Lema\^itre law connecting galaxy distances with redshifts --- something
which was also previously addressed\footnote{Building upon Fath (1909),
  Slipher (1915, 1917), and Campbell \& Paddock (1918) who had identified
  NGC~4151 as a spiral nebula with a redshifted velocity of 940 km s$^{-1}$,
  Lundmark (1925) referred to these redshifts as ``Campbell shifts'', now
  known as ``Hubble redshifts''.  Although nowadays very much in the shadow of
  Hubble (1929), Lundmark (1925, p.867) wrote that ``A rather definite
  correlation is shown between apparent dimensions and radial velocity, in the
  sense that the smaller and presumably more distant spirals have the higher
  space-velocity'', and he presented a second-order distance-velocity relation
  which effectively reduces to Hubble's (1929) first-order distance-velocity
  relation upon setting $m=0$.}  by Lundmark (1925).

As Block et al.\ (2004) reminded us, it was actually Sir James Jeans (1928)
who came up with the Y-shaped diagram that encapsulated this early-to-late
type galaxy sequence that was later turned sideways and became known as the
Hubble tuning fork (Figure~\ref{fig:tuning_fork}).  Rather than a single
linear sequence, the two arms of the Y-shaped figure were invoked to capture
the spiral nebulae that Curtis (1918) had noted come both with and without
bars.  In Hubble's (1936) version of this Y-shaped diagram, he added the
lenticular (S0) galaxy type from Reynolds (1925) --- the so-called ``armless
spiral'' galaxies, if ever there was an oxymoron --- at the nexus point.
Including this previously hypothetical galaxy type from Jeans (1919a,b),
Hubble (1936) expanded upon Hubble (1926) by expressing an awareness of the
continuity of forms among the entire (luminous) galaxy population, as opposed
to an elliptical branch versus two spiral branches.  However, the variation in
ETG morphology was not yet fully appreciated at this time.

\section{Revisions, adaptions, enhancements} \label{Sec_Mods}

Having a better understanding of the tuning fork diagram's origin, we are
well-placed to now appreciate and understand the revisions and re-expressions
that it has experienced over the years.

\subsection{de Vaucouleurs classification volume}

The tuning fork diagram has had a remarkable longevity, and witnessed a
continual stream of proposed upgrades and complementary classification
schemes.  For example, Shapley \& Paraskevopoulos (1940) introduced the Sd
galaxy type, and de~Vaucouleurs (1959a) introduced the Magellanic-like Sm
galaxy type, and also the Im galaxy type for Irregular (Irr) galaxies related
to the Small Magellanic Cloud, which included both barred and unbarred
versions.  Around this time, Holmberg (1958) introduced the subdivisions,
i.e.\ the Sab, Sbc galaxy types.

Working with Hubble's (unpublished) notes written prior to his death in 1953,
de Vaucouleurs (1959a) points out that one of the main revisions that Hubble
wanted to make to the galaxy classification system was the introduction of
three groups of barred lenticular galaxies, defined largely according to the
contrast and strength of the bar.  This was captured by de Vaucouleurs (1959a,
1959b, see his Figures~9 and 11), which introduced the weak/anemic bar class
(AB) for both spiral and lenticular galaxies, slotting in between galaxies
with a clear bar (designated by the letter B) and those without (designated by
the letter A).  de Vaucouleurs (1959a) also replaced the phrase, ``normal
spiral'' with ``ordinary spiral'', to reflect the fact that comparable numbers
of barred and non-barred spiral galaxies were observed, and thus one type was
not more normal than the other.

These additions can be thought of as converting the tuning fork into a
trident, with a new middle prong for the AB types having weak bars.  However,
de~Vaucouleurs (1959a,b) stepped things up yet another level, and introduced
not just three parallel sequences (A, AB, B) for the S0 and spiral galaxies,
but also an orthogonal third axis to the galaxy classification scheme in order
to designate whether the spiral arms\footnote{Danver (1942) predicted that a
  classification of just the arm patterns may one day be needed, effectively
  echoing Reynolds (1927) criticism of the simplified system of Hubble (1926)
  which overlooked the different character of the spiral arms. Four decades
  later, Elmegreen \& Elmegreen (1982) brought in a fine microscope and
  introduced 12 distinct arm classes.} originate from a ring or not (r=ring,
rs=mixed, s=s-shaped spiral with no ring: see Hodge 1966).  This culminated in
a three-dimensional classification volume, which was further embellished with
additional morphological details, which can be appreciated from the extensive
classification types in the Third Reference Catalog of Bright Galaxies (de
Vaucouleurs et al.\ 1991).

Arguably, the three-dimensional classification volume of de Vaucouleurs
(1959a) departed too far from the simplicity of the two-dimensional tuning
fork.  Indeed, there were three stages, from early- to late-type, just within
the lenticular galaxy class (S0$^-$, S0$^0$, S0$^+$); this was in addition to
the E$^+$ stage bridging the E and S0 galaxies, and the S0/a stage bridging
the S0 and Sa galaxies.  The various bells and whistles added by de
Vaucouleurs resulted in a classification scheme approaching the complexity of
that introduced by W.Herschel (Curtis 1933, see p.919) or the 5-number
sequence introduced by his son J.Herschel (Lundmark 1927, p.19).  This level
of complexity reflected a recognition that ETGs are not simple,
single-component systems.  However, in spite of de Vaucouleurs classification
volume, many research papers continue to flounder by treating ETGs as though
they are single component systems.  This may in some instances arise from
recourse to the simplicity of the tuning fork.

\subsection{Sandage's Hubble Atlas}

Sandage, Hubble's former PhD student who had taken possession of Hubble's
notes and loaned them to de~Vaucouleurs in the mid 1950s, used these to
produce his own ``Hubble Atlas of Galaxies'' (Sandage 1961; see also Sandage
1975).  The properties of the spiral arms --- such as their winding angle, the
degree of condensation --- rather than the apparent 
bulge-to-total ($B/T$) flux ratio, were now made the primary criteria in
establishing the spiral morphological types.  In a second cautious distinction
from Hubble (1926), Sandage (1961, page 6) suggested that the
Aitken-Jeans-Lundmark-Hubble sequence was an evolutionary one, such that
galaxies started out as ``late-type spirals'' with young stars that evolved into
``early-type spirals'' with old stars that subsequently morphed into ``early-type
galaxies'', but he clearly noted that it was an unproven idea.  His motivation
came from the known trends involving the stellar populations (e.g.\ 
Morgan \& Mayall 1957; Morgan 1958).

As noted by Sandage (1961), disc galaxies of a given Hubble type had a range
of $B/T$ flux ratios.  This is because, as just noted, the first criteria used
in Sandage's classification scheme was the openness of the spiral arms,
followed by the resolution of the arms into stars.  Progressing from ETGs
which do not have spiral arms, to the Sa galaxies having tightly wound spiral
arms with pitch angles of a few to $\sim$10 degrees, to the late-type spirals
with open arms and pitch angles of up to $\sim$40 degrees, there is a
sequence.  The existence of some S0 galaxies with a low $B/T$ flux ratio is
still consistent with the picture that the S0 galaxies are a transition class
between the E and Sa galaxy types.  However, this breaks down somewhat in
schemes using the $B/T$ flux ratio as the primary diagnostic while the nature
of the spiral arms takes second-place.

No doubt in part due to the greater ease in estimating $B/T$ flux ratios than
spiral arm winding angles and resolution (which can be impeded by the
disc inclination to our line-of-sight and image depth), modern classifications
have a tendency to follow Hubble's and de Vaucouleurs' ordering of the
criteria and align themselves more with the $B/T$ flux ratio than with the spiral
arms.  Today, each spiral galaxy type, and the S0 galaxy type (which can
contain rings, i.e.\ the most tightly wound of all spiral arms), do still
possess a range of $B/T$ flux ratios (e.g.\ Freeman 1970; Boroson 1981; Kent
1985; Kodaira et al.\ 1986; Simien \& de Vaucouleurs 1986; Graham \& Worley
2008; Laurikainen et al.\ 2010).  This is in part because different
classifiers have used alternating primary and secondary criteria to classify
galaxies.

Sandage (1961; see also Figure~75 in Bigay 1963) also had three subgroups of
S0 galaxy, depending on the degree and location of dust lanes, plus three
subtypes of barred S0 galaxy, depending on the nature of the bar.  Indeed,
Sandage, de Vaucouleurs, and Hubble, considered bars to be a primary criteria
in their classification schemes.

\subsubsection{Bars}

While Jeans (1902) explored the stability and balance (in the early-stage of spherical
nebulae) between the outward 
pressure of a hot gas cloud and the inward force of gravity, Safronov (1960;
see also Gurevich \& Lebedinsky 1950) additionally considered the shear forces
arising from the differential (Keplerian) rotation within the discs of both
proto-planetary clouds and the Milky 
Way.  Toomre (1964), in collaboration with Agris Kalnajs, expanded this to the
discs of external galaxies.  In addition to Hunter (1963), who explored disc
stability using oblate spheroidal coordinates and solutions to Laplace's
equation, Safronov and Toomre established idealised stability criteria which
could lead to the coagulation of discs and also the formation of bars (Hohl 1971).
While subsequent work including dark matter halos offered a stabilising
environment for discs, bars do still form from these global disc instabilities
in which the circular disc orbits become elongated to form the bar
(e.g.\ Combes \& Sanders 1981; Combes \& Elmegreen 1993; Sellwood \& Wilkinson
1993; Athanassoula 2003).  Bars can also experience instabilities of their
own, resulting in a buckling both within and above the disc plane (Combes et
al.\ 1990; Athanassoula 2005; Saha et al.\ 2018) to produce X/boxy/(peanut
shell)-shaped ``pseudobulge'' structures or ``barlenses'' which were added to the expanded
galaxy classification scheme of Buta et al.\ (2010).

Bars themselves are also a recognised driving force in secular galaxy evolution.  Given how
the dynamical (in)stability of the disc can be seen through the appearance of
bars, the tuning fork diagram, plus the extension to include weak bars,
reflects an important physical process.  Bars may also offer insight into the
triaxiality, central density and rotation of the dark matter halos thought to
surround galaxies (e.g.\ Debattista \& Sellwood 2000; Athanassoula 2002, 2003;
Berentzen et al.\ 2006).  It is therefore considered desirable to retain the
bar strength in a schematic of galaxy classification.

\subsection{Yerkes system}

The Yerkes system of galaxy classification used the radial concentration of
light, as suggested by Lundmark (1925), plus optical spectra to trace
the stellar populations (Humason et al.\ 1956; Morgan \& Mayall 1957; Morgan
1958, 1959; Morgan \& Osterbrock 1969).  
For luminous galaxies, the subjectively-defined concentration parameter set 
(denoted: a, af, f, fg, g, gk, k) 
helped distinguish between the relative prominence of the bulge and disc, and
it reportedly 
correlated better with the central stellar population than the Hubble 
morphological-type did.\footnote{While the Yerkes system proposed replacing 
the Sa, Sab,
Sb, ...\ sequence with the radial concentration of the light, 
it retained the four broad ``form families'' (Elliptical, Spiral,
Barred Spiral, and Irregular) as a secondary parameter
and added four more: dusty elliptical,
nucleated, low surface brightness, and rotationally-symmetric but without
clear elliptical or spiral structure. In addition, the 
axis-ratio was used to define a second 
secondary parameter, referred to as the ``inclination class'', which ranged 
from 1 (spherical) to 7 (elongated).}  The 
S\'ersic (1963) index $n$ (see Graham \& Driver 2005 for a review of the
$R^{1/n}$ model) 
was also designed to quantify the varying bulge-to-disc flux ratio, and Graham et
al.\ (2001) showed that the galaxy S\'ersic index is monotonically related to
the galaxy's mean concentration index (Okamura et al.\ 1984).  However, unlike
with the S\'ersic index, this objectively-defined 
concentration parameter is rather sensitive to the
image exposure depth (Graham et al.\ 2001; see also Povi\'c et al.\ 2015).
Furthermore, dwarf ETGs have the same concentrations as late-type galaxies
(LTGs), and thus a single concentration parameter does not define the
ETG-to-LTG sequence if one wishes to include dwarf galaxies.  While the
inclusion of galaxy type, or rather ``form family'', within the Yerkes system
resolves this ambiguity, it also highlights the value in retaining 
a morphological descriptor.

The central, (optical luminosity)-weighted spectra provided mean age estimates
along the galaxy sequence, offering hope of unlocking the formation history of
galaxies.  It revealed an abundance of early-type stars in late-type galaxies,
and late-type stars in early-type galaxies.  Although helpful, as can be
inferred from van den Bergh (1975, his p.60-61), mass-weighted ages, or
star-formation rate histories of the separate bulge and disc components, are
more desirable and less prone to mis-direction. 
Studying spectra from the bulges of spiral galaxies, 
MacArthur et al.\ (2009) found that both early- and late-type spiral galaxies
possess old mass-weighted ages, with less than one-quarter 
of the stellar mass arising from young stars.  
Variants of the Lundmark-Yerkes system are still used today, as seen in
the concentration-colour plane for bright galaxies (e.g.\ Driver et
al.\ 2006). 

The cD galaxies (Matthews et al.\ 1964; Morgan \& Lesh 1965) of the Yerkes
system are galaxies with three-dimensional shrouds, known as halos or
envelopes, which belong more to the host galaxy's cluster than to the cluster's
central cD galaxy. These galaxies are typically ETGs, and not some
new kind of galaxy.

\subsection{The van den Bergh trident}

The David Dunlap Observatory (DDO) classification system for late-type
galaxies (van den Bergh 1960a,b,c) introduced luminosity classes based on the
correlation between the absolute magnitude of spiral galaxies and the degree
to which their spiral arms are developed.  It was later expanded to include
dwarf galaxies (van den Bergh 1966).  The DDO system was substantially
re-worked by van den Bergh (1976), who now based it on disc galaxies with
either: strong spiral arms, i.e.\ spiral galaxies (S); weak/anemic spiral
arms, i.e.\ S0/a galaxies (A); or no spiral arms (S0 galaxies).  This produced
the three prongs of a trident whose handle was the E0-E6 sequence.  Attention
to the arm strength was favoured over the bar strength, and the principle axis
of the disc galaxy sequence now reflected the bulge-to-total flux ratio (as in
Hubble 1926) rather than the nature, i.e.\ the winding angle and
condensations, of the spiral arms (as in the original DDO system and Sandage 1961).  As such,
the presence of bars was now overlooked in not only the ETGs but also the
LTGs.  However, the three prongs served to capture a continuity in spiral arm
contrast that mimicked the varying gas content and luminosity-weighted stellar
ages.  Despite the substantial changes, van den Bergh referred to this as the
Revised David Dunlap Observatory (RDDO) system.

The motivation behind this change was to reflect the galaxy formation scenario
of Spitzer \& Baade (1951) rather than Jeans (1919a).  Building on (i) the
idea in Spitzer \& Baade (1951) that S0 galaxies are gas-stripped spiral
galaxies, which have subsequently lost their spiral pattern, due to the
environment of a galaxy cluster, and (ii) inspired by Sandage, Freeman \&
Stokes (1970, see also Rood \& Baum 1967) who remarked that S0 and spiral
galaxies have the same distribution of axial ratios, van den Bergh (1976)
embraced the notion of S0 galaxies as a parallel sequence to spiral galaxies.
This was in contrast to the notion that they are a bridging sequence between
elliptical and spiral galaxies.  However, such an origin from spiral galaxies,
or at least a universal origin, was later ruled out because many S0 galaxies
are more luminous than spiral galaxies (e.g.\ Burstein et al.\ 2005), and
therefore alternative formation mechanisms are required for S0 galaxies.
Furthermore, S0 galaxies do not just reside in clusters.  Therefore, at least
some are not stripped spiral galaxies.  Indeed, ETGs with the same disc-like
photometric and kinematic properties as those in galaxy clusters are known to
exist in isolation (Janz et al.\ 2017; Graham et al.\ 2017).

Aside from the above issues negating the motivation for this trident, another
obvious setback with the proposal for a true ``parallel sequence'' (among the
prongs) based on the bulge-to-disc flux ratio is that the spiral galaxies are
not structurally similar to ETGs with discs.  That is, spiral galaxies do not
have the high bulge-to-disc ratios of some ETGs, and in general ETGs galaxies
do not have the low bulge-to-disc ratios of many late-type galaxies
(e.g.\ Graham \& Worley 2008).  For example, ETGs can have discs fully
contained within their spheroidal structures, 
they are neither E nor S0 galaxies, while many 
late-type galaxies can be bulgeless but an abundance of such S0 counterparts
are not known.  The markedly different histogram of bulge-to-total flux ratios
between ETGs with discs and the spiral morphological sequence was shown by
Freeman (1970, his Figure~9).  Graham et al.\ (2016, their Figure~7) provides
an expanded representation of this histogram sequence, 
better depicting the bulge-to-total flux
ratio for early- and late-type galaxies. It shows the distinction between the
bulge-dominated 
ES galaxies, S0 galaxies with both significant large-scale discs and low
bulge-to-total flux ratios, and the tail in the distribution to bulgeless
spiral galaxies.  This highlights the incomplete nature of the prongs in the
trident, i.e.\ that there is not a parallel sequence but rather only partial
overlap which superficially appears substantial if one includes (with no
number-density weighting) the underpopulated-wings of the bulge-to-total distribution from
each morphological type.

Graham \& Worley (2008) show that S0 galaxies with low bulge-to-total ($B/T$)
flux ratios are relatively rare, yet they should not be {\it if} they formed
from the gas-stripping of spiral galaxies.  This is because late-type spiral
galaxies are less massive than the early-type spiral galaxies --- remember the
DDO system (van den Bergh (1960c) --- and have lower stellar densities in
their discs (e.g.\ Graham 2001, his Figure~3).  Therefore, they should be
easier to convert into gas-free, and eventually spiral-less, S0 galaxies than
it is to transform the early-type spiral galaxies with their higher $B/T$ flux
ratios.  

Given all of the above factors, there is a preference for (full) parallel
sequences involving bar strength, rather tnan (partial) parallel sequences
involving spiral arm strength.

Rather than building S0 galaxies by stripping the discs of spiral galaxies, it
may be that S0 galaxies are formed by accretion and the building of discs
(e.g.\ Young et al.\ 2008; Kannappan et al.\ 2009; Wei et al.\ 2010; 
Alatalo et al.\ 2013; Moffett
2014; Graham et al.\ 2015, and references therein; Mahajan et al.\ 2018).
This latter process is truncated in galaxy clusters, perhaps contributing to the
relatively low numbers of spiral galaxies in clusters when compared to the
field population.

\subsubsection{Further developments}

Modelling bars as separate components, and using the S\'ersic (1963) $R^{1/n}$
model for the bulge component, the decomposition of ETG images by Laurikainen
et al.\ (2010) confirmed that most ordinary ETGs are disc-dominated, and it revealed a
wide range of $B/T$ flux ratios that was also observed by Krajnovi\'c et
al.\ (2013).  This range of $B/T$ flux ratios motivated Cappellari et
al.\ (2011) to adopt the trident classification of van den Bergh (1976),
although they drew his trident slightly differently, re-aligning the location
of the trident's handle (E0 to E5 in their scheme) with the trident's outer
``lenticular galaxy'' prong, and declaring a new paradigm for ETGs which they
called the ATLAS$^{3D}$ comb.  In addition, the degree of rotational support
now supplanted the $B/T$ ratio along the primary axis of the ATLAS$^{3D}$ comb
(Cappellari et al.\ 2011, their Figure~2), thereby embracing the scheme
discussed by Bender (1988), Capaccioli \& Caon (1992), Kormendy \& Bender
(1996) and others based on a galaxy's angular momentum.

While Cappellari et al.\ (2011, their Figure 1) note that the spiral galaxies
do not actually have the same high $B/T$ flux ratios as the ETGs, they did use
NGC~4452 --- whose thin inner disc/bar is prominent at optical wavelengths due
to its edge-on orientation --- to suggest that ETGs may have $B/T$ ratios as
low as late-type spiral galaxies. However, due to its edge-on orientation, it is
unknown what type of galaxy NGC~4452 is, as with IC~335 and the Spindle Galaxy
NGC~5866.\footnote{Measurements of the health (strength and thinness) of such
  discs, reflecting a lack of heating, i.e.\ movement of stars out of the disc
  plane, or a lack of significant accretion (e.g.\ Brook et al.\ 2004), 
are impossible to make for most galaxies and thus not incorporated
  into morphological classification schemes.}  
 Simulations show that spiral
galaxies, when viewed edge-on, can look like NGC~4452 (e.g.\ Valentini et
al.\ 2017, their Figure~6).  Moreover, such {\it potential} S0 galaxies are 
not only rare, but one may ask, Where are the face-on ETGs with small or no
bulge?\footnote{Kormendy \& Bender (2009) provide a 5-component fit to the major-axis
of NGC~4452, with the bar and barlens inverted, and report a very small $B/T$
ratio of $\sim$0.02.  A similar edge-on galaxy is NGC~4762 (PGC~043733: van
den Bergh 1976; Baillard et al.\ 2011; Jarrett et al.\ 2003), which has had
its (geometric mean)-axis light profile modelled by Sahu et al.\ (2019) and
has a $B/T$ flux ratio of 0.08, on par with Sc galaxies (Graham \& Worley
2008, their Table~4). For comparison, Sdm/Sm galaxies tend to have ratios 2 to
3 times smaller, or no bulge (e.g.\ Walcher et al.\ 2005).}

Cappellari et al.\ (2011) suggested that the anemic spiral galaxies --- midway
between the S0 and strong spiral galaxies --- have small
amounts of gas and a passively evolving stellar population, and they
associated these with the ``red spirals'' (e.g.\ Masters et al.\ 2010).
However, Cortese (2012) subsequently revealed that these red spiral galaxies
have star formation rates on par with ordinary spiral galaxies, undermining
the suggestion that the ``red spirals'' are a bridging population in terms of
reduced amounts of star formation activity, and issuing a warning for
classifications schemes based on optical colour.  Furthermore, the ETGs (with
$M_B < -14$~mag) also have a range of colours, and a range of masses, voiding
the notion that either colour or mass might monotonically vary across the
trident/comb.  The blue, low-mass ETGs illustrate this point well (Driver et
al.\ 2006; Lee et al.\ 2006; Mei et al.\ 2006; Deng et al.\ 2009; Schawinski
et al.\ 2009; Kaviraj et al.\ 2011; George \& Zingade 2015; Graham et
al.\ 2017).  Nonetheless, for non-dwarf galaxies, there are some trends and
partial parallel sequences within the trident/comb, in
which for a given 
$B/T$ flux ratio, a range of disc galaxy types may be found. 

Kormendy \& Bender (2012) amended their modified tuning fork from Kormendy \&
Bender (1996, which included three modifications taken from de Vaucouleurs
1959a, and which is discussed next), to present a somewhat similar scheme to
van den Bergh's trident and the ATLAS$^{3D}$ comb.  However, Kormendy \&
Bender (2012) placed the dwarf ``elliptical'' galaxies --- which they called
dwarf spheroidal galaxies in order to distance them from ordinary ETGs on the
left of the diagram --- next to the Irregular Magellanic type galaxies at the
far right of the diagram.  Graham (2019) explains in some detail the false
dichotomy between dwarf and ordinary ETGs that was based on bends at $M_B
\approx -18$ mag in structural parameter diagrams involving the arbitrary 50\%
radius, $R_{\rm e}$, and the associated surface brightness terms.  Use of
radii containing different percentages of a galaxy's light reveals how the
location of the bends change by more than 3~mag, and thus the bend mid-point and
the separation of dwarf and ordinary ETGs has nothing to do with galaxy
formation processes but instead depends on the arbitrary definition of galaxy
radii.  Related calls for a division, or continuity, among the ETG population 
are also addressed in Graham (2019).

\subsection{Isophotal shape}\label{Sec_Iso}

Using a Fourier analysis to quantify the departure of galaxy isophotes from
pure ellipses, in an effort to detect photometrically weak discs via the presence
of pointy/lenticular-shaped isophotes, as previously done by Carter (1978, 1987) and
Cawson (1983), Bender (1988) wrote that ``It is most curious that there is
little evidence for significant morphological difference between rapidly and
slowly rotating ellipticals''. The answer to this curiosity is that the
observed (on the plane of the sky) ellipticity of a disc varies with
$\cos(i)$, while the kinematic contribution from the disc is measured
perpendicular to the plane-of-the-sky, parallel to our line-of-sight, and
varies with $\sin(i)$, where the disc's angle of inclination $i$ equals 0 degrees 
when seen face-on.  Tables of $\cos(i)$ and $\sin(i)$ will immediately reveal
why the presence of a disc is more apparent in the kinematics than in an
image.  Indeed, this justifiably led Cappellari et al.\ (2011) to prefer the
use of kinematic measures rather than isophotal shapes for their ATLAS$^{3D}$
comb.

Bender (1988; see also Nieto et al.\ 1988) investigated the relation between
the mean isophotal shape, i.e.\ boxy or discy, 
and the ratio of the inner (typically $\lesssim 1 
R_{\rm e}$) stellar rotation and mean velocity dispersion, $V / \sigma$.
Bender (1988) suggested that elliptical galaxies can be separated into two
classes: those that are more closely related to the S0 galaxies, such that
they are rotationally flattened with weak discs and have discy isophotes; and those
with boxy isophotes that are dynamically supported by anisotropy in their velocity
dispersion. Kormendy \& Bender (1996) subsequently  visualised this through their modified
representation of the elliptical galaxies in the tuning fork, now based on
boxy-to-discy isophotes rather than ellipticity. Although, they noted a number
of complications, and two others are mentioned here.

Kormendy \& Bender (1996) reported that the measured velocity anisotropy
correlated with the mean isophotal shape and advocated for using the isophotal
shape to represent the elliptical galaxies in
the Jeans-Hubble tuning fork.  However, Figure~2 in Kormendy \& Bender (1996),
which plots isophotal shape versus $V / \sigma$, shows that galaxies with boxy
isophotes have no correlation with $V / \sigma$. This may be a consequence
of sampling the dynamics of both pure elliptical galaxies and massive S0
galaxies.  
Lenticular galaxies can have massive merger-built bulges --- with
partially-depleted cores created by coalescing supermassive black holes
(Begelman et al.\ 1980) ---
and large-scale discs (e.g.\ Dullo \& Graham 2014).  Their
$B_4$ profiles\footnote{$B_4$ is the amplitude of the fourth-order Fourier
  cosine term used to describe variations from purely elliptical isophotes
  (e.g.\ Carter 1978; Ciambur 2015).} vary with radius, changing from boxy to
discy as the radius increases.  An average inner isophotal shape does not capture the
two-component nature of these galaxies.  

Depending on the orientation of the
disc to our line-of-sight, ES galaxies can have high ellipticities and
discy-shaped isophotes at small radii --- along with considerable rotational
support within the apertures typically used to measure the kinematics 
 --- which then transition to low-ellipticities and
elliptical, or even boxy-shaped, isophotes at large radii (e.g.\ Nieto et
al.\ 1991, their section~4.2).  An edge-on example
of this is LEDA~074886 (Graham et al.\ 2012; see also NGC~4638 in Ferrarese et
al.\ 2006, their Figure~13), while a face-on example is LEDA~2108986 (CG~611:
Graham et al.\ 2017).  At least in the latter case, a fundamental formation
mechanism is evident through the accretion of this ES galaxy's disc.  In those
ES galaxies where the kinematical measurements have been obtained at radii beyond the
intermediate-scale disc, the rotation is observed to fall away (e.g.\ Arnold
et al.\ 2014: Foster et al.\ 2016; Bellstedt et al.\ 2017; Rawlings et
al.\ 2019).

A long string of papers have advocated for a divide between ETGs with boxy
isophotes and a partially depleted core versus those with discy isophotes and
either no depleted core or a central excess.\footnote{Jerjen \& Binggeli
  (1997) may have been the first to consider additional nuclear components in
  ETGs as defined relative to the galaxy's outer S\'ersic profile; Graham et
  al.\ (2003) were the first to define and model the central deficit of light
  in ETGs relative to the inward extrapolation of the spheroid's outer
  S\'ersic profile.}  However, such campaigns, which have effectively treated
ETGs as single component systems, via a single $B_4$ parameter, have 
overlooked both the ES galaxy population bridging the E-to-S0 galaxies, and the massive
lenticular galaxies noted above. 
The use of a single discy or boxy parameter, or indeed a single
ellipticity parameter, can be misleading for ETGs. Classification based on the
shape of the radial isophotal {\it profile}, 
rather than a single mean isophotal shape, would help when the discs
are relatively edge-on. However, this approach will fail when the discs are
relatively face-on.  Classification by kinematics offers advantages here, but
it is more expensive in terms of telescope time.  For ETGs, a classification
scheme recognising the continuum of disc extent 
relative to the extent of the spheroid (see Simien \&
de Vaucouleurs 1986; Capaccioli et al.\ 1988) --- obtained from an analysis
and decomposition of the image --- would be preferable to a single isophotal
shape parameter which neither adequately captures the two-component
(bulge/disc) nature of ETGs nor represents the radially varying kinematics and
isophotal shapes in ES galaxies.

\subsection{Quantitative footings: CAS space}

The galaxy classification scheme of Lundmark (1925, 1927) involved the
concentration\footnote{``The ratio of the nuclear light to the light of the
  outer portions''.}, and compressibility\footnote{Lundmark simply described
  this as ``Different degrees of                                                            
  compressibility towards [the] galaxy centre''.}, of galaxies within each of his
four galaxy classes (ETG, LTG, Magellanic, Peculiar) discussed in the
excellent review, at that time, of galaxies by Lundmark (1927).  Lundmark's
classes reflected the symmetry, or lack thereof, of the galaxies.  The
classification scheme of Hubble (1926, 1936) was also related to the 
concentration of the 
later type galaxies --- as traced through the apparent bulge-to-disc flux
ratio --- in combination with the apparent smoothness of the ETGs versus the
increasing condensations and clumpiness seen in the LTGs.  Hubble (1936)
has additionally noted how the galaxy's optical colour correlated with the
morphological type. 

Although the ellipticity parameter in Hubble's classification scheme can be
more dependent on the viewing angle than the intrinsic
major-to-minor axis of a galaxy, it offered an objective quantity rather
than a subjective description.  Not surprisingly, building on the Yerkes
system, several objectively defined concentration parameters were later
introduced, e.g.\ Fraser (1972, see also de Vaucouleurs 1977) and Okamura et
al.\ (1984, later popularised by Doi et al.\ 1992, 1993 and subsequently Abraham et
al.\ 1994).  
As was recognised by Okamura et al.\ (1984), and
later quantified by Graham et al.\ (2001), their mean concentration parameter is a
rather strong function of the exposure depth, and, as such, it does not simply
measure the intrinsic concentration of a galaxy's light.  Trujillo et
al.\ (2001) therefore introduced a new concentration parameter to minimise the influence
of the galaxy exposure depth, and Okamura et al.\ (1984) had suggested that
additional parameters, such as the mean surface brightness, could be
introduced to try and reduce this observational bias and limitation of the
mean concentration parameter.

Expanding on this push towards a parameterised version of the Jeans-Lundmark-Hubble
galaxy classification scheme resulted in the introduction of (a)symmetry
parameters to measure galaxy peculiarities and non-symmetries in the arms
(e.g., Lindblad \& Elvius 1952; Schade et al.\ 1995; Rix \& Zaritsky 1995;
Abraham et al.\ 1996; Conselice 1997) and the introduction of smoothness
parameters to measure star-forming clumpiness or image ``bumpiness'' (e.g.,
Isserstedt \& Schindler 1986; Takamiya 1999; Conselice 2003; Blakeslee et
al.\ 2006).  Collectively, Concentration, Asymmetry, and Smoothness, have been
used to produce a three-dimensional CAS galaxy classification scheme
(Conselice 2003).  Although not without its issues (e.g.\ Neichel et
al.\ 2008; Huertas-Company et al.\ 2015), including image depth, spatial
resolution, and galaxy orientation, it offers a helpful quantification of the
properties used by Lundmark and Hubble.  

In contrast with de Vaucouleurs' three-dimensional classification volume ---
with its axes of what is effectively bulge/disc ratio, bar strength and ring
pattern --- the CAS scheme reduces the emphasis on the disc instability that
leads to the creation of bars (and the ensuing ansae and rings), in favour of
better capturing merger activity and star-formation.  This is increasingly
important when probing higher redshifts (e.g.\ Griffiths et al.\ 1994;
Driver et al.\ 1995; Giavalisco et al.\ 1996; Mortlock et al.\ 2013), where
the Jeans-Lundmark-Hubble classification becomes less applicable due to the
rising fraction of galaxies with Irregular structures.  By increasing the
emphasis on condensations, non-smoothness, asymmetries, and irregularities ---
which were already features of the Lundmark and Hubble classification schemes
-- the CAS scheme may better facilitate the identification of the physical
processes which shaped galaxies, albeit at the expense of a helpful taxonomy
for identifying distinct structural components of evolved galaxies.

Classifications based on various kinds of interactions have also been advanced
(e.g.\ Vorontsov-Vel'Yaminov 1959), and an array of special symbols were used
in the extensive Morphological Catalog of Galaxies (MCG: Vorontsov-Vel'Yaminov
\& Arkhipova 1962-1974). Cross correlation with the classification of de
Vaucouleurs (1959a,b) and Sandage (1961) can be found in Vorontsov-Vel'Yaminov
\& Noskova (1968).  Not surprisingly, colour has also been used (e.g.\ Conselice
1997), although the presence of colour gradients in galaxies, and different
colours for different components, plus the colour-magnitude relation for ETGs,
and the above mentioned star-formation in ``red spirals'', complicates the
derivation and limits the usefulness of a single optical colour parameter for
galaxies.

Despite its merits, the CAS space still leaves want for a simple two-dimensional
graphic which captures the primary morphology of established galaxies.

\begin{figure*}
\begin{center}
\includegraphics[trim=0.5cm 10.9cm 0.7cm 10.5cm, width=1.0\textwidth]{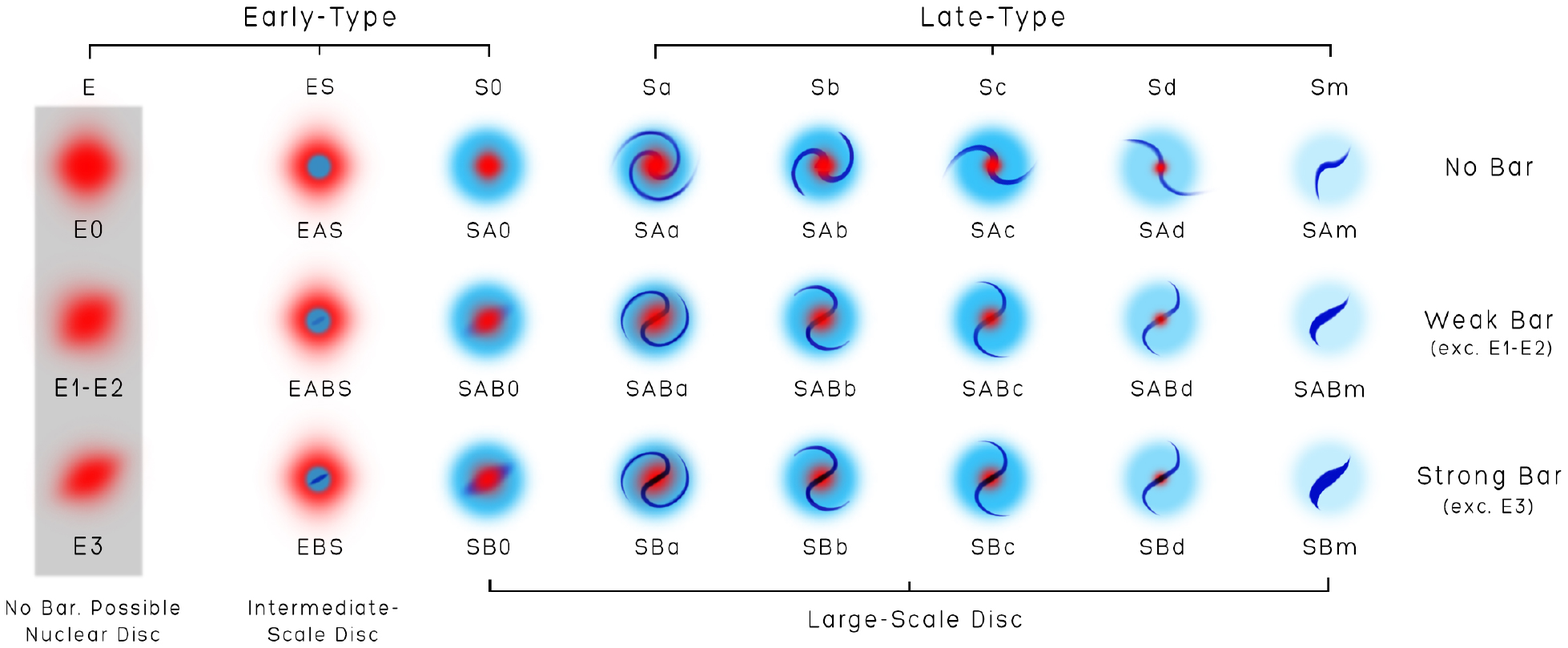}
\caption{A two-dimensional grid of morphological types designed to provide a
  finer representation for the ETG population by better highlighting their
  discs and bars.  There are direct parallels with the Jeans-Hubble tuning
  fork and de Vaucouleurs' (1959a) three-dimensional galaxy classification
  volume, especially if one considers de Vaucouleurs' E$^+$ galaxies to be the
  ES galaxies of Liller (1966). The relative prominence of the bulge and disc
  can be seen through the ETG section of the grid. The (largely) discless
E0 to E3 sequence is slightly set aside in order to distance it from the bar
designations on the right.}
\label{fig:grid}
\end{center}
\end{figure*}

\section{A galaxy morphology classification grid}\label{Sec_Grid}

As noted earlier, for a long time ``nebulae'' were not known to contain spiral
patterns, and therefore Laplace's (1796) nebular hypothesis did not generate
spiral structures.  However, after spiral nebulae were first observed by
Parson in Rosse (1850), Aitken (1906) modified the Laplace model in an effort to
explain these structures, and Jeans (1919a) subsequently embraced this scenario.  When
Hubble (1926) presented his classification scheme based upon the Jeans galaxy
evolution sequence and the barred/non-barred spiral sequence of Reynolds
(1920), it did not include lenticular S0 galaxies, but they were later added
in Hubble (1936) once they had become more commonly known.  When de
Vaucouleurs (1959a) and Sandage (1961) presented their revised galaxy
classification schemes, they did not mention ETGs with intermediate-scale
discs.  Such galaxies were not
highlighted until Liller (1966).  Although a spate of papers from 1988 to 1993
drew attention to this type of galaxy, to date, galaxy morphology diagrams
remain somewhat lacking when it comes to this type of galaxy. 

At different radii, 
individual ETGs can be both boxy and discy, a slow rotator and a fast rotator,
in addition to possibly possessing a bar within a disc of differing radial extent.  To 
help address this, Graham et al.\ (2017) presented a schematic of the typical
bulge-to-disc flux ratio versus galaxy morphological type, revealing how the ES galaxies fit
in.  Figure~\ref{fig:grid} presents a complementary grid to better highlight
the discs and bars in ETGs.  
There is an inherent usefulness in summarising the continuity of forms of
today's galaxies into an easy-to-grasp (two-dimensional) diagram. Theories and
simulations need to address when and how these forms arose (e.g.\ Fiacconi et
al.\ 2015; Genel et al.\ 2015; Remus et al.\ 2015).  Despite its critics, from
Reynolds (1927) to Abraham (1996) and Conselice (2003), the
Jeans-Lundmark-Hubble scheme went a long way to achieving this summary.  By
more fully populating the two-dimensional space (of form and bar strength) and
switching from a tuning fork to a grid, greater information about the ETG
population  can be incorporated.

Figure~\ref{fig:grid} is something of a compromise between the
simplicity of the tuning fork and the complexity of de Vaucouleurs'
classification volume.  The grid has effectively collapsed de~Vaucouleurs
(1959a) (S0$^-$, S0$^0$, S0$^+$) series back down to the single S0 galaxy type
included by Hubble (1936), and treated de~Vaucouleurs (1959a) E$^+$ stage as
though they are the ES type galaxies from Liller (1966).  As such, the
classifications shown in Figure~\ref{fig:grid} are familiar --- albeit with the
exception of the non-barred (EAS), weak-barred (EABS) and strong barred (EBS)
continuity within the ES population --- and the classifications can additionally be
embellished with informative higher-order labels, as done by de Vaucouleurs
(1959a,b), Buta et al.\ (2010), and Laurikainen et al.\ (2011) to denote the
presence of components, such as rings and nucleation, that are increasingly modelled and quantified in modern
galaxy decomposition work (e.g.\ Davis et al.\ 2019; Sahu et al.\ 2019).

As with ordinary ETGs, dwarf ETGs can also have spheroid/disc/bar/etc.\ components.
They both occupy the left hand side of the grid, encompassing systems with a range
of $B/T$ flux ratios but without a spiral density wave.  

Two old caveats are noted.  The perceived ellipticity of a Maclaurin (1742) ellipsoid or
Jacobi (1834) triaxial structure can change with one's line-of-sight to the
structure.  Furthermore, due to the presence of intermediate- and large-scale
discs, the ellipticity often changes with radius for the same line-of-sight to
the galaxy.  This should be kept in mind when viewing the handles of tuning
forks, tridents, and the elliptical galaxy edge of the grid shown in
Figure~\ref{fig:grid}.  Second, a galaxy's bulge-to-disc flux ratio and spiral
arm pitch angle measurements can vary from author-to-author (due to
methodologies, bandpass, measurement errors), and the image concentration
depends on the observer's definition, exposure depth and measurement
technique.  Similarly, Lahav et al.\ (1995) reminded us that the assigned
morphological type also varies from author-to-author.  As such, there is a
certain level of scatter along the principle-axis of the grid, whether it be
defined primarily by the spiral arms, bulge-to-disc flux ratio, or the radial
concentration of the light if excluding the dwarf galaxies.

While numerous additional schemes to morphologically classify galaxies have
their merits (e.g.\ van den Bergh 1960c; Lekshmi et al.\ 2003; Lotz et
al.\ 2004; Yamauchi et al.\ 2005; Conselice 2006; Scarlata et al.\ 2007; Vika
et al.\ 2015; Ferrari et al.\ 2015; Selim \& Abd El Aziz 2017), it is hoped
that the simplicity of, and continuity within, the grid shown in
Figure~\ref{fig:grid} will be a helpful expansion of the tuning fork, serving
as a reminder that some ETGs also have bars, and that they possess a range of
disc sizes.  The grid does not attempt to be all-encompassing.  Not included
are dwarf
spheroidal galaxies (van den Bergh 1959), early- and late-type ultra
diffuse galaxies with low surface brightnesses and diameters of $\sim$10 kpc
(Sandage \& Binggeli 1984), rare (less than a few percent at $z=0$)
Irregular galaxies, nor interacting galaxies (Vorontsov-Ve1yaminov 1977). 
Rather than tack the latter two classes on at 
the end of the grid, the Irregular and Peculiar galaxies are considered here
to be unrelaxed disturbed 
systems which are yet to settle into one of the types shown on the grid.

\section{Discussion}\label{Sec_Disc}

It has often been repeated over the last century that the merit of a galaxy
classification scheme can be measured by its ability to track evolutionary
pathways.  As noted earlier, the changing, luminosity-weighted, mean age of
the stellar population along the Jeans-Lundmark-Hubble sequence was routinely
heralded as a success of this classification scheme.  We now know that this
suggestion by Sandage (1961), and others before him, that this age-sequence might reflect
an evolutionary pathway from right to left along the tuning fork, was not
correct.  However, a galaxy classification scheme has an additional benefit:
it creates familiarity with the morphology, structure, and components that
galaxies are comprised of.

It is hoped that the galaxy morphology classification grid shown in Figure~\ref{fig:grid}
will raise awareness as to the presence and range of disc sizes in early-type
galaxies.  For example, Savorgnan \& Graham (2016b) reported on how a lack of
awareness of intermediate-scale discs led to claims of over-massive black
holes in ES galaxies reported to have unusually high $M_{\rm bh}/M_{\rm
  bulge}$ ratios.  However, some of the galaxies had previously been modelled with
large-scale discs, resulting in an underestimation of the bulge light.

To give another example,
over the past decade, many clever
studies have tried to understand how practically all compact massive spheroids at
$z\sim2.0\pm0.5$ could grow their 3D spheroid to become elliptical galaxies
(e.g.\ Hopkins et al.\ 2009; Hilz et al.\ 2013; Oogi \& Habe 2013, and
references therein).  However, the similar 
sizes, masses, densities, S\'ersic indices of local ($z=0$) bulges and high-$z$
compact massive galaxies suggests an alternate solution to the fate of some 
of the high-$z$ compact massive galaxies which may have accreted/formed
a disc rather than turned into an elliptical galaxy (Graham et al.\ 2015). 
For lenticular galaxies, their 2D discs can dramatically increase
the size of the galaxy beyond the size of their 3D bulge.  The stellar density
within lenticular {\it galaxy} half light radii are less than the stellar density
within the half light radii of their {\it bulge} components because bulges
naturally pack in more stars by filling the 3D volume while discs only occupy
a thickened plane.  As shown in Graham (2013, his Figure~1), the density of
bulges is higher than the density of ETGs of the same mass, except at the
high-mass end where the ETGs are 3D spheroidal stellar systems.

\subsection{Formation paths}

ETGs are often said to constitute the endpoint of galaxy evolution, such that
they have exhausted their gas supply from which stars are formed, making them
``red and dead''. However, this picture is somewhat limited given the
``down-sizing'' nature to galaxy growth, in which lower-mass field galaxies
form from smaller over-densities in the early Universe, and take longer to
accrete their neighbouring gas supply.  Given their smaller masses,
significant fractions of material can still swoop in and continue to build
these galaxies today.  Indeed, the dwarf early-type galaxy CG~611 has young spiral arms in
its intermediate-scale stellar disc that is undergoing growth via gas
accretion in counter rotation to the stellar disc (Graham et al.\ 2017).  The
past growth of discs via accretion and minor mergers is also evident in more
massive ETGs, with counter rotation seen in, for example, NGC~7007 and
NGC~4550 (Dettmar et al.\ 1990; Rix et al.\ 1992). Therefore, the popular notion that gas poor
ETGs with discs evolved from gas rich spiral galaxies is probably not the full
picture.  Rather, the accretional growth of discs building ES and S0 galaxies 
may be a fundamental pathway for galaxy growth over the past ten billion
years. 

Perhaps a key limitation of the classification of galaxies in their ``ground
state'', and thus a preference for the interaction types of the MCG
(Vorontsov-Vel'Yaminov \& Arkhipova 1962-1974) and the asymmetry and
clumpiness indices, is that these may better reveal a progression or sequence
of galaxies traversing the ``excited states'' to reach the ``ground state''
configuration.  The hope of using the Jeans-Hubble tuning fork as a tool for
understanding galaxy evolution, in a manner akin to how the
Hertzsprung-Russell (HR) diagram (Rosenberg 1910; Hertzsprung 1911; Russell
1914; Nielsen 1964) is used for understanding stellar evolution, may have
therefore been partly hamstrung from the get go.  This was suspected at least
as far back as Shapley \& Paraskevopoulos (1940) who wrote, in regard to the
1000+ classified galaxies (e.g.\ Shapley \& Ames 1932), that: ``The [rare]
unusual forms, whether chaotically irregular ... or merely peculiar variations
on the usual types ... [are] of uncommon interest and perhaps of special
importance in the study of galactic structures and development.''  As such,
the largely overlooked classification scheme for galaxy interactions
(Vorontsov-Vel'Yaminov 1959) may be where one should look for evolutionary
pathways.  In a sense, while the HR diagram sampled stars in different phases
of their evolution, perhaps the Jeans-Hubble diagram, with its emphasis on big evolved
galaxies, in essence samples galaxies at a single (evolved) epoch.  Much of the
evolution may effectively be hidden in the connection between the now ($z=0$)
rare Irregular galaxies  and the common structured galaxies, at odds with the
(previously) expected notion of evolution between the structured galaxies.  This is not to say that
the latter does not occur.  Indeed, disc accretion has likely converted some
elliptical galaxies into ellicular and lenticular galaxies (Graham et
al.\ 2015), which  would represent limited evolution from left to right
along a section of the the Jeans-Lundmark-Hubble sequence, as Jeans first
speculated, but due to the accretion of external material rather than due to
cooling, contraction, and spinning-up of an isolated nebula.

The grid presented in Figure~\ref{fig:grid} recognises the existence of bars
in ETGs, as does de Vaucouleurs (1959a) three-dimensional classification
volume and the Hubble-Sandage classification scheme (Sandage 1961), and it
also captures the broader range of sizes of the spiral-less discs in ETGs that
was first recognised by Liller (1966).  It is noted, however, that ETGs, and 
in particular ES and S0 
galaxies, are not limited to being three-component bulge$+$disc$+$bar systems
(e.g.\ Cheng et al.\ 2011).  
Building upon the practice of simultaneous multi-component decompositions
(e.g.\ de Jong 1996; Prieto et al.\ 1997, 2001), ETGs not only have bars
within discs (e.g.\ Friedli \& Martinet 1993; Laurikainen et al.\ 2005), but
they may also possess barlenses / (peanut shell)-shaped pseudobulges (Combes
et al.\ 1990; Athanassoula 2005; Athanassoula et al.\ 2015; Laurikainen et
al.\ 2011, 2018; Saha et al.\ 2018), ansae (e.g.\ Martinez-Valpuesta et
al.\ 2007; Saha et al.\ 2018), nuclear-rings, bar-rings and outer-rings (Theys
\& Spiegel 1976; de Vaucouleurs et al.\ 1991; Michard \& Marchal 1993; Buta \&
Combes 1996; Buta 2011; Mirtadjieva \& Nuritdinov 2016; Buta 2017), and
nuclear stellar discs (e.g.\ Scorza \& Bender 1995; Scorza et al.\ 1998;
Morelli et al.\ 2004; Balcells et al.\ 2007) which likely form a continuum
with the intermediate-scale discs of the ES galaxies 
and the large-scale discs of the S0 galaxies.  This
variety of components found within ETGs can be appreciated from the
multi-component decompositions by Buta et al.\ (2010), L\"asker et
al.\ (2014), Savorgnan \& Graham (2016a), Davis et al.\ (2019), and Sahu et
al.\ (2019).

Detailed information about the spiral arms in LTGs (e.g.\ Elmegreen \&
Elmegreen 1982) may require higher-dimensions than offered by a
two-dimensional schematic, although see Block \& Puerari (1999) and Seigar et
al.\ (2005) in regard to use of spiral arm number and pitch angle in the
near-infrared.  As noted earlier, such higher-order information from more
complex schemes can always be annotated to a primary classification, as done
with the de Vaucouleurs volume.

\subsection{Kinematic classifications}

Although this study has focussed on the morphological classification of
galaxies, it is insightful to review some of the relevant developments
pursuing a kinematic classification.  As noted previously, the prevalence of
discs in ETGs has taken a remarkably long time to be realised.  Indeed, most,
if not all, relaxed E4 to E6/E7 galaxies are misclassified S0 galaxies (Liller
1966; Gorbachev 1970; Michard 1984; Capaccioli et al.\ 1990; van den Bergh
1990).  Although many papers have pointed to the existence and abundance
of discs in ETGs (e.g.\ Bertola \& Capaccioli 1975; Davies et al.\ 1983;
Capaccioli 1987; Carter 1987; Nieto et al.\ 1988, 1991; Rix \& White 1990;
Sandage \& Bedke 1994; Scorza \& Bender 1995; Graham et al.\ 1998; Rix et
al.\ 1999; Krajnovi\'c et al.\ 2013; Scott et al.\ 2014; Bassett et al.\ 2017), in some quarters
they are still not widely recognised, nor fully appreciated.  However, this
seems set to change with the increased amount of kinematic data becoming
available.

After reporting, for several years, on the missed discs in ETGs, Capaccioli \&
Caon (1992) wrote that ``It was in fact possible to catalog ellipticals into
two kinematical families: one dominated by random motions and made up of the
most luminous galaxies, and another containing fainter objects characterized
by a fair balance between random and ordered motions, just as in [the] bulges
of S0 and S galaxies'' and that the ETGs with discs describe ``a sequence
primarily described by the value of the specific angular
momentum''.\footnote{This is somewhat reminiscent of the work by Alexander
  (1852), in which the angular momentum of the collapsing nebula was the
  determining factor in the production of elliptical or spiral nebulae; see
  also Shaya \& Tully (1984).}  Indeed, this is the basis of the kinematic
classification which the SAURON project (Bacon et al.\ 2001; de Zeeuw et
al.\ 2002) used to separate ETGs into slow or fast rotators (Emsellem et
al.\ 2007).

Of course, what the SAURON project were in essence doing was re-labelling the E
galaxies as ``slow rotators'' (SR), and the S0 and ES galaxies as ``fast
rotators'' (FR).  The strength in their approach was the ease at which
they could detect discs in their kinematic data (including line-of-sight
velocity $V$, and velocity dispersion $\sigma$) due to the $\sin(i)$, rather
than $\cos(i)$, dependence on the inclination of the disc, as noted in
Section~\ref{Sec_Iso}.  Furthermore, analysis of the shape of the absorption
lines, and their deviations from a Gaussian, offered additional means for probing
the embedded discs (Krajnovi\'c et al.\ 2008).  
Additional observations using the
SAURON integral-field spectrograph on the William Herschel Telescope
spectroscopically 
confirmed an 
abundance of discs in ETGs (Emsellem et al.\ 2011; Cappellari et al.\ 2011).  The
ATLAS$^{3D}$ team subsequently explored the bulge$+$disc nature of these ETGs
through bulge$+$disc decompositions (Krajnovi\'c et al.\ 2013), and the
flattened nature of these discs, as opposed to rotating ellipsoids, was shown
through their study of the galaxy's axial ratios (Weijmans et al.\ 2014),
building on Freeman (1970) and Sandage et al.\ (1970).

Considering ETGs to be single-component ellipsoids that can be flattened by
either rotation or anisotropy in their velocity dispersion, Binney (1976,
1978, 1985; see also de Zeeuw \& Franx 1991) established a dividing line
between fast and slow rotators in the ellipticity-($V_{\rm max}/\sigma_0$)
diagram.  Capellari et al.\ (2007) used this diagram to argue that the fast
rotators are nearly oblate and contain disc-like components.  Emsellem et
al.\ (2007, see their Figure~5) introduced a clever variation by deriving the
aperture ``spin'' parameter, $\lambda_R$, which replaced $V_{\rm
  max}/\sigma_0$, and where the subscript $R$ denoted the radius of the
aperture.  This $\epsilon$--$\lambda_R$ diagram has been used to classify
galaxies, including ETGs, as either an FR or SR.

However, this FR/SR classification suffers from one of the problems that
undermined the use of isophotes to classify ETGs as either boxy or discy.
Specifically, unlike with spiral galaxies, for ETGs the presence of an inner
disc does not ensure the presence of a disc at large radii.  The FR versus SR
designation has therefore propagated the tendency for one to overlook the ES
galaxies which are both fast rotators at small radii and slow rotators at
large radii.\footnote{Emsellem et al.\ (2007, see their Figure~2) were
  aware of this population, but the FR/SR convention that they used
  does not encapsulate it.}  To help rectify this situation, Graham et al.\ (2017)
suggested a further variation to the ellipticity-kinematic diagrams such that
galaxies are not represented by a single aperture value but rather annular
information is used to show the radially-varying behavior of their kinematics
and ellipticity.  Bellstedt et al.\ (2017) further revealed how incorporating the {\it
  radial} kinematic and photometric information aids in the identification of
intermediate-scale discs, and counter-rotating discs, with galaxies moving
along ``radial tracks'' in the modified spin-ellipticity
($\epsilon(R)$--$\lambda(R)$) diagram. 

Expanding upon Krajnovi\'c et al.\ (2006), van de Sande et al.\ (2017) also
kinematically classified galaxies, introducing five classes using the velocity
dispersion, rotation, skewness and kurtosis of the stellar absorption lines,
albeit using a single measure for each of these quantities within one
effective half light radius, $R_{\rm e}$.  When the ``integral field
spectrograph'' data allows it, it may be additionally fruitful to represent the array of
galaxy-wide information using more than one aperture radius (e.g.\ Chung \&
Bureau 2002).  Based on the 2D kinematic maps, Krajnovi\'c et al.\ (2011)
divided the regular and non-regular rotators into seven subgroups: 
featureless; low-rotation; kinematic twist; kinematically decoupled cores; 
counter rotating cores; double maxima in the radial velocity profile (2M galaxies); double
maxima in the velocity dispersion map, i.e.\ 2 $\sigma$ peaks.  Most
recently, Rawlings et al.\ (2019) have attempted to condense this 2D information
via a suite of template radial tracks in the modified spin-ellipticity
diagram for seven different morphological-dynamical types of galaxy, including
the ES galaxies with intermediate-scale discs, the ``2 $\sigma$ peak''
galaxies, plus E, S0, early- and late-type spiral galaxies, and barred spiral
galaxies.

\subsection{The future with Big Data sets}

The Universe is big.  The observable portion of the Universe may contain in
excess of $10^{11}$ galaxies, notably higher than the, at the time impressive,
figure of nearly one million galaxies circa 1930.  The dramatic increase in
galaxy sample sizes since the early 1900s has resulted in an explosion of data
and catalogs that all too often is not matched by the human resources required
to fully analyse it.  This inability to classify every galaxy using the ways
of the past, i.e.\ visual inspection, could potentially undermine the current
classification scheme.  Indeed, there have been calls to replace galaxy
morphological types with quantitative metrics that machines can quickly
calculate.  This final subsection serves to acknowledge the merits of such a
suggestion based on robust non-degenerate metrics, 
and to briefly note how the community is moving forward on this front
while still embracing key elements seen in both the tuning fork and the galaxy
morphology classification grid. 

Visual classification by a dedicated team of professionals
(e.g.\ Vorontsov-Vel'Yaminov \& Arkhipova 1974; de Vaucouleurs et al.\ 1991;
Driver et al.\ 2006; Fukugita et al.\ 2007; Buta et al.\ 2010; Nair \& Abraham
2010; Ann et al.\ 2015; Kartaltepe et al.\ 2015) has its limitations at around
$10^4$ galaxies.  As such, we have seen the birth of Citizen Scientist
projects, with Galaxy Zoo recruiting up to $\sim$10$^5$ budding and amateur
astronomers who, collectively, can visually classify far more galaxies
(e.g.\ Lintott et al.\ 2008; Willett et al.\ 2013; Dickinson et al.\ 2018).
Furthermore, new
branches of the galactic community are evolving through the exploration and
application of software capable of semi-automatic galaxy classification on a
large scale (e.g.\ Sedmak \& Lamas 1981; Huertas-Company et al.\ 2015).
Examples of this Machine Learning (de la Calleja \& Fuentes 2004; Bazell \&
Miller 2005; Shamir 2009; Barchi et al.\ 2019) include Support Vector
Machines, statistical learning methods such as Classification Trees with
Random Forest (CTRF) and Neural Networks (Huertas-Company et al.\ 2008;
Diaz-Hernandez et al.\ 2016; Sreejith et al.\ 2018; Sultanova 2018), and
enhanced brain storm optimization techniques (Ibrahim et al.\ 2018).  
Beck et al.\ (2018) describe how the preferably-large training
sets for machine learning can effectively be provided by citizen scientists.
Spectral
(e.g.\ Morgan \& Mayall 1957; S\'anchez Almeida et al.\ 2010; Wang et
al.\ 2018) and kinematic (e.g.\ Brosche 1970; Wakamatsu 1976;
Baiesi-Pillastrini 1987; Wiegert \& English 2014; Hung et al.\ 2015; van de
Ven et al.\ 2016; Kalinova et al.\ 2017; van de Sande et al.\ 2017; Rawlings
et al.\ 2019) information is additionally being folded into the mix,
supplementing and complementing the morphological classification process.

\section*{acknowledgements}

It is a pleasure to thank the Department of Astronomy at The University of
Florida, Gainesville, USA, for hosting me as a part of Swinburne University of
Technology's Research Sabbatical Scheme.  I am additionally grateful to
James Josephides who converted my rough sketches into the Figures shown
herein.  This research was supported by Australian Research Council funding
through grant DP110103509.

\end{document}